\documentclass{aa}
\usepackage{graphicx}
\usepackage{natbib}
\usepackage{txfonts}
\begin{document}
   \title{H-atom bombardment of CO$_2$, HCOOH and CH$_3$CHO containing
   ices}

   \subtitle{}

   \author{S.~E. Bisschop \and G.~W. Fuchs \and E.~F. van Dishoeck
     \and H. Linnartz}

   \offprints{S.~E. Bisschop, bisschop@strw.leidenuniv.nl}

   \institute{Raymond and Beverly Sackler Laboratory for Astrophysics,
     Leiden Observatory, Leiden University, P.O. Box 9513, 2300 RA
     Leiden, the Netherlands }

   \date{Received; accepted}

 
  \abstract
  {Hydrogenation reactions are expected to be among the most important
    surface reactions on interstellar ices. However, solid state
    astrochemical laboratory data on reactions of H-atoms with common
    interstellar ice constituents are largely lacking.}
  {The goal of our laboratory work is to determine whether and how
    carbon dioxide (CO$_2$), formic acid (HCOOH) and acetaldehyde
    (CH$_3$CHO) react with H-atoms in the solid state at low
    temperatures and to derive reaction rates and production yields.}
  {Pure CO$_2$, HCOOH and CH$_3$CHO interstellar ice analogues are
    bombarded by H-atoms in an ultra-high vacuum experiment. The
    experimental conditions are varied systematically. The ices are
    monitored by reflection absorption infrared spectroscopy and the
    reaction products are detected in the gas phase through
    temperature programmed desorption. These techniques are used to
    determine the resulting destruction and formation yields as well
    as the corresponding reaction rates.}
  {Within the sensitivity of our set-up we conclude that H-atom
    bombardment of pure CO$_2$ and HCOOH ice does not result in
    detectable reaction products. The upper limits on the reaction
    rates are $\leq$7$\times$10$^{-17}$~cm$^2$~s$^{-1}$ which make it
    unlikely that these species play a major role in the formation of
    more complex organics in interstellar ices due to reactions with
    H-atoms. In contrast, CH$_3$CHO does react with H-atoms. At most
    20\% is hydrogenated to ethanol (C$_2$H$_5$OH) and a second
    reaction route leads to the break-up of the C--C bond to form
    solid state CH$_4$ ($\sim$20\%) as well as H$_2$CO and CH$_3$OH
    (15--50\%). The methane production yield is expected to be equal
    to the summed yield of H$_2$CO and CH$_3$OH and therefore CH$_4$
    most likely evaporates partly after formation due to the high
    exothermicity of the reaction. The reaction rates for CH$_3$CHO
    destruction depend on ice temperature and not on ice
    thickness. The results are discussed in an astrophysical context.}

   \authorrunning{Bisschop et al.}  \titlerunning{H-atom
   bombardment of CO$_2$, HCOOH and CH$_3$CHO ices} \maketitle
%

\section{Introduction}\label{intro}

It is generally assumed that at the low temperatures in interstellar
clouds, thermal hydrogenation of molecules on icy grain surfaces is
the main mechanism to form more complex saturated species (Tielens \&
Charnley 1997). This is due to the relatively high abundance of
H-atoms in the interstellar medium as well as their high mobility even
on cold grains.  However, most of the reactions in the proposed
reaction schemes have not yet been measured experimentally.
\citet{hiraoka1994,hiraoka2002}, \citet{watanabe2004},
\citet{hidaka2004} and \citet{fuchs2007} have studied reactions of
thermal H-atoms with CO ice in the laboratory, and shown that H$_2$CO,
and at higher fluxes also CH$_3$OH, are readily formed at temperatures
as low as 12 K. It thus seems likely that other species will also be
able to react with H-atoms to form saturated grain-surface
products. These species may be the starting point for an even more
complex chemistry that occurs at higher temperatures or by energetic
processing due to UV or cosmic rays in the ice. Eventually, the ices
will evaporate when heated by a protostar, leading to the complex
organics seen in hot cores \citep[e.g.,][]{blake1987,ikeda2001}. The
aim of this paper is to study the reactivity of a number of
astrophysically relevant molecules with H-atoms in interstellar ice
analogues at low temperatures to test the proposed thermal
hydrogenation reaction scheme and to characterize which products are
formed and which mechanism is involved.

Interstellar ices contain both simple and complex species
\citep[see][]{ehrenfreund1999}. The most abundant ice molecules are
H$_2$O, CO and CO$_2$, which have very strong vibrational modes. The
spectroscopic identification of other less abundant ices, such as
HCOOH and CH$_3$CHO studied here, relies on weaker bands i.e., the OH
and CH bending modes, $\nu_{\rm B}$(OH/CH), of HCOOH at 7.25~$\mu$m
and the CH$_3$ deformation, $\nu_{\rm D}$(CH$_3$), of CH$_3$CHO at
7.41~$\mu$m. Solid state abundances of HCOOH are 1--5\% in both low
and high mass star forming regions with respect to H$_2$O
\citep{schutte1997,schutte1999,gibb2004,boogert2004}. The detection of
CH$_3$CHO is less certain, but abundances up to 10\% have been
reported \citep{gibb2004,keane2001a}.

\begin{figure}\centering
\includegraphics[width=8cm]{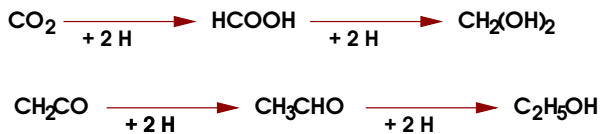}
\caption{Potential reaction routes for hydrogenation of CO$_2$ and
  CH$_2$CO ice. Only stable products are
  shown.}\label{reaction-scheme}
\end{figure}

The specific species studied in this paper are CO$_2$, HCOOH and
CH$_3$CHO, molecules that take a central place in inter- and
circumstellar hydrogenation reaction schemes
(Fig.~\ref{reaction-scheme}). CO$_2$, HCOOH and CH$_2$(OH)$_2$ differ
only in their number of hydrogen atoms. It is therefore possible that
they are related through successive hydrogenation reactions. Previous
laboratory experiments of H$_2$ and CO$_2$ have resulted in the
formation of HCOOH on ruthenium surfaces \citep{ogo2006}, but this
reaction may have been mediated by the catalytic surface. 

Another series of organics that are thought to be linked through
successive hydrogenation are CH$_2$CO, CH$_3$CHO and C$_2$H$_5$OH
(Fig.~\ref{reaction-scheme}). Ethanol is indeed detected in warm gas
phase environments in star-forming regions but CH$_2$CO and CH$_3$CHO
are found mostly in colder gas \citep{ikeda2001,bisschop2007a}. This
may be either due to very efficient conversion of CH$_3$CHO into
C$_2$H$_5$OH or because the latter species may be formed through
another route. In particular, astronomical observations show a
constant CH$_3$OH/C$_2$H$_5$OH ratio which indicates that these two
species are chemically linked \citep{bisschop2007a}.

H-addition reactions in astrophysically relevant ices have been
studied previously through UV photolysis experiments, where the
hydrogen atoms are produced by dissociation of a suitable precursor
molecule, often H$_2$O ice
\citep[e.g.,][]{ewing1960,milligan1964,milligan1971,ijzendoorn1983,allamandola1988,gerakines2000,moore2001,wu2002}. Although
these experiments give a useful indication whether certain
hydrogenation reactions may or may not occur, their results cannot be
compared directly with those obtained from the laboratory studies
mentioned above, nor can they be used to quantitatively test reaction
schemes such as those in Fig.~\ref{reaction-scheme} (see \S~\ref{plab}
for details).

This paper is organized as follows: \S~\ref{expt_meth} explains the
experimental method, \S~\ref{sec_analysis} focuses on the data
reduction and analysis, \S~\ref{co2_sec}--\ref{ch3cho_sec} discuss
the results, derived reaction rates and chemical physical mechanisms
for hydrogenation reactions with HCOOH, CO$_2$ and CH$_3$CHO,
\S~\ref{disc} presents the astrophysical implications and finally
\S~\ref{sum} summarizes the main conclusions of this study.

\section{Experiments}
\label{expt_meth}

\subsection{Our experiment}\label{our_expt}
The experiments are performed using a new ultra-high vacuum set-up
that comprises a main chamber and an atomic line unit. The details of
the operation and performance of the set-up are described by
\citet{fuchs2007}. The main chamber contains a gold coated copper
substrate (2.5$\times$2.5~cm$^2$) that is mounted on top of the cold
finger of a He cryostat. Temperatures can be varied between 12 and
300~K with 0.5~K precision using a Lakeshore 340 temperature control
unit and are monitored with two thermocouples (0.07\% Au in Fe versus
chromel) that are mounted on the substrate face and close to the
heater element. The typical pressure in the main chamber during
operation is better than 5$\times$10$^{-10}$~mbar.

Pure ices of $^{13}$C$^{18}$O$_2$ (97\% purity, Icon),
$^{12}$C$^{18}$O$_2$ (97\% purity, Icon), HCOOH (98\% purity,
J.~T. Baker) and CH$_3$CHO (99\% purity, Aldrich) as well as mixed
ices of $^{13}$C$^{18}$O$_2$ with H$_2$O (deionized) and CO (99.997\%
purity, Praxair) are studied (see Table~\ref{expt-hcooh}, for an
overview of the mixture ratios, ice thicknesses and ice
temperatures). The two isotopologues of CO$_2$ are used to distinguish
between atmospheric CO$_2$ and solid CO$_2$ processed by H-atoms. The
ices are grown at 45$^0$ with a flow of 1.0$\times$10$^{-7}$~mbar~s$^{-1}$
where 1.3$\times$10$^{-6}$~mbar~s$^{-1}$ corresponds to 1~monolayer (ML)
s$^{-1}$. The temperatures of the ices range from 12 to 20~K and their
thicknesses are chosen between 8 and 60~ML.

H-atoms are produced in a well-studied thermal-cracking device
\citep{tschersich1998,tschersich2000}. The dissociation rate and
resulting H-atom flux depend on the temperature and pressure which are
both kept constant during a single experiment. The temperature of the
heated Tungsten filament, $T_{\rm W}$ is $\sim$2300~K in all
experiments and the H+H$_2$ flow through the capillary in the atom
line is either 1.0$\times$10$^{-4}$ or
1.0$\times$10$^{-5}$~mbar~s$^{-1}$. For the latter pressure the
calculated dissociation rate, $\alpha_{\rm dis}$, in the atomic line
is 0.45. The atoms that exit the source are hot. Before the H-atoms
enter the main chamber, they pass through a quartz pipe and
equilibrate to $\sim$300~K. The minimum number of collisions of
H-atoms with the quartz pipe is 4 due to the nose-shaped form of the
pipe. Since H-atoms are thermalized with the surface after only 2--3
collisions, it is expected that most H-atoms will have temperatures
equal to the quartz pipe of 300~K. Due to collisions with the walls of
the pipe and with each other, a fraction of the atoms recombines to
H$_2$. The effective dissociation fraction, $\alpha_{\rm dis}^{\rm
  eff}$, and H-atom flux on the sample surface are therefore
calculated to be lower than given by \citet{tschersich2000}, i.e.,
0.13 and 5.0$\times$10$^{14}$~cm$^{-2}$~s$^{-1}$, or 0.20 and
7.8$\times$10$^{13}$~cm$^{-2}$~s$^{-1}$ for the chosen flow-rates of
1.0$\times$10$^{-4}$ and 1.0$\times$10$^{-5}$~mbar~s$^{-1}$,
respectively \citep{fuchs2007}. Note also that, the absolute number of
H-atoms on the surface at a given time is not equal to the surface
flux, because processes such as scattering and recombination take
place on the surface. Theoretical simulations show that a steady-state
H-atom coverage of 5.0$\times$10$^{14}$~cm$^{-2}$~s$^{-1}$ with an
error of a factor 2 is quickly reached in this regime independent of
the exact H-atom flux (Cuppen, private communication). A more
extensive discussion of the derivation of these steady-state numbers
is given by \citet{fuchs2007}. The time and H-fluence, i.e., the total
number of atoms~cm$^{-2}$ integrated over time in each experiment are
listed in Table~\ref{expt-hcooh}. At the temperatures and fluxes used
in our experiment, the substrate surface will be covered with H$_2$ in
a few seconds. Since H$_2$ molecules do not stick to other H$_2$
molecules, the maximum coverage with H$_2$ will only be a few
monolayers. H-atoms therefore have to diffuse through the cold H$_2$
layer before reaching the ice and will be completely thermalized with
the surface at the moment they encounter the ice sample. Experiments
with only H$_2$ molecules have been performed for comparison by
setting the source temperature $T_{\rm W}$ to $\sim$600~K as an
additional check to confirm that reactions are due to H-atoms and not
to H$_2$ molecules (see Table~\ref{expt-hcooh}).

The ices are monitored by Reflection Absorption Infrared Spectroscopy
(RAIRS) using a Fourier Transform Infrared Spectrometer, covering
4000--700~cm$^{-1}$ with a spectral resolution of 4~cm$^{-1}$. The
infrared path length is the same for all experiments. Typically 512
scans are co-added. An experiment starts with a background RAIR scan
and subsequently an ice is deposited onto the substrate
surface. Another RAIR spectrum is taken after deposition to determine
the initial number of molecules in the ice. An additional background
spectrum is recorded afterward such that subsequently recorded spectra
yield difference spectra between ices before and after H-atom
bombardment. The next step in the experiments is the continuous H-atom
bombardment of the ice during which a RAIR scan is taken every
10~minutes. After 3~hrs the experiment is stopped and a Temperature
Programmed Desorption (TPD) spectrum is obtained using a quadrupole
mass spectrometer. The ramp speed is 2~K~min$^{-1}$ and is continued
until the temperature reaches 200~K. Control experiments with pure
ices of CO$_2$, C$_2$H$_5$OH, CH$_4$ and CH$_3$OH are studied to
determine RAIR band strengths and to calibrate the production yields
that are measured using the mass spectrometer (see
\S~\ref{yieldsec}). In those cases RAIR spectra are taken right after
deposition and subsequently TPD spectra are recorded as described here
for the other experiments.

\begin{table}
  \caption{Summary of all H-atom bombardment experiments for pure
    CO$_2$, HCOOH and CH$_3$CHO ices, as well as the mixed
    morphologies. The total time that the ices are exposed to H-atoms
    are indicated with $t$ and the temperature of the tungsten
    filament with $T_{\rm W}$. A value $T_{\rm W}$ of 2300~K refers to
    an experiment in which H$_2$ is dissociated and of 600~K to
    control experiments of ices bombarded with
    H$_2$-molecules.}\label{expt-hcooh}
\begin{center}
\begin{tabular}{lllll}
  \hline
  \hline
  Thickness & $T_{\rm ice}$ & $t$ & $T_{\rm W}$ & H-fluence\\
  (ML)      & (K)           & (min)         & (K) & (molecules cm$^{-2}$) \\
  \hline

\multicolumn{5}{l}{CO$_2$}\\
\hline
15        & 12.5          & 180           & 2300 & 5.4(18)\\
15        & 14.5          & 180           & 2300 & 5.4(18)\\
15        & 14.5          & 180           & 600 & -- \\
\hline
\multicolumn{5}{l}{CO$_2$:H$_2$O}\\
\hline
15$^a$        & 14.5          & 180           & 2300 & 5.4(18)\\
15$^b$        & 14.5          & 180           & 2300 & 5.4(18)\\
15$^c$        & 14.5          & 180           & 2300 & 5.4(18)\\  
15$^a$        & 14.5          & 120           & 2300 & 3.6(18)\\
15$^a$        & 14.5          & 180           & 600  & --\\
\hline 
\multicolumn{5}{l}{CO:$^{13}$C$^{18}$O$_2$}\\
\hline
30$^d$  & 14.6 & 180 & 2300 & 5.4(18)\\
15$^d$  & 14.6 & 180 & 2300 & 5.4(18)\\
30$^d$  & 14.6 & 180 & 2300 & 5.4(18)\\
45$^d$  & 14.6 & 180 & 2300 & 5.4(18)\\
15$^e$  & 14.6 & 180 & 2300 & 5.4(18)\\
30$^e$  & 14.6 & 180 & 2300 & 5.4(18)\\
\hline
\multicolumn{5}{l}{HCOOH}\\
\hline
20        & 12.5          & 240       & 2300 & 7.2(18)\\
20        & 40.0          & 240       & 2300 & 7.2(18)\\
20        & 12.5          & 240       & 600  & ---\\
\hline
\multicolumn{5}{l}{HCOOH:H$_2$O$^f$}\\
\hline
40        & 12.5          & 180           & 2300 & 5.4(18)\\
\hline
\multicolumn{5}{l}{CH$_3$CHO}\\
\hline
16.2        & 14.5       & 180        & 2300 & 5.4(18)\\ 
7.8         & 14.5       & 180        & 2300 & 5.4(18)\\
11.4        & 14.5       & 180        & 2300 & 5.4(18)\\
13.5        & 14.5       & 180        & 2300 & 5.4(18)\\
18.8        & 14.5       & 180        & 2300 & 5.4(18)\\
21.2        & 14.5       & 180        & 2300 & 5.4(18)\\
22.1        & 14.5       & 180        & 2300 & 5.4(18)\\
56.0        & 14.5       & 180        & 2300 & 5.4(18)\\
45.8        & 14.5       & 180        & 2300 & 5.4(18)\\
11.4        & 12.4       & 180        & 2300 & 5.4(18)\\
11.3        & 17.4       & 180        & 2300 & 5.4(18)\\
11.2        & 19.3       & 180        & 2300 & 5.4(18)\\
11.7        & 14.5       & 180        & 600  & ---\\
\hline
\end{tabular}
\end{center}

$^a$39:61\% $^{13}$C$^{18}$O$_2$:H$_2$O, $^b$22:78\% $^{13}$C$^{18}$O$_2$:H$_2$O, $^c$48:52\% $^{13}$C$^{18}$O$_2$:H$_2$O, $^d$45:55\% $^{13}$C$^{18}$O$_2$:CO, $^e$80:20\% $^{13}$C$^{18}$O$_2$:CO, $^f$HCOOH:H$_2$O 20:80\%.
\end{table}

\subsection{Comparison with other hydrogenation experiments}
\label{plab}

A large number of photolysis experiments of astrophysically relevant
ices exist where H-atoms are produced through photo dissociation of
H$_2$O or other precursors
\citep[e.g.,][]{ewing1960,milligan1964,milligan1971,ijzendoorn1983,allamandola1988,gerakines2000,moore2001,wu2002}. These
give useful information on potential hydrogenation reactions schemes,
but do not give specific information about reaction rates (see for
example \S~\ref{co2_disc} \& \ref{discon}). Also the question whether
thermal hydrogenation reactions can be responsible for newly formed
species is not answered, for several reasons. First, the hydrogen
atoms resulting from photolysis are produced {\it in situ} inside the
ice with an excess energy of several eV; such atoms can travel
significant distances through the ice \citep[e.g.,][]{andersson2006}
and a reaction may take place before thermalization is achieved. Thus,
activation energy barriers can be overcome, in contrast to thermal
hydrogenation reactions where this is less probable. Second, although
dilution in an inert Ar matrix can stabilize the H-atoms, the H-atom
flux on the reactants remains poorly characterized. Third, other
reactive products such as OH are also formed by photolysis of H$_2$O
which makes it hard to discriminate the different effects. It is not
possible to study pure ices through this method since H$_2$O or
another precursor is always needed to provide a source of
H-atoms. Finally, the photolysis experiments reported so far have been
carried out under high vacuum conditions (typically $10^{-7}$~mbar) in
which several monolayers of background gases (mostly H$_2$O) are
accreted in less than a minute, providing additional molecules that
can be photolyzed during the experiments. This may affect the
outcome. In contrast, our experiments and the previously mentioned
surface science experiments on CO hydrogenation are performed under
ultra-high vacuum conditions (typically $10^{-10}$~mbar) in which less
than a monolayer of background gas (mostly H$_2$) is accreted during
the time-scale of the experiments (a few hours). Here the H-atoms are
formed by a microwave source or by thermal cracking and are
thermalized to room temperature or less before striking the ice
surface, rather than being produced inside the ice.

\section{Data analysis}
\label{sec_analysis}
\subsection{RAIR analysis}
\label{rair_an}
Different frequency ranges are selected for baseline subtraction that
depend on the species under study. Fourth order polynomial baselines
are fitted to the recorded RAIR spectrum. Additionally, local third
order polynomial baselines are subtracted around the features of
interest to accurately determine the integrated absorption. The
frequency ranges are given per ice morphology in Table~\ref{int}. From
the integrated intensity of the infrared bands, the column density of
species X in the ice is calculated through a modified Lambert-Beer
equation:

\begin{equation}\label{column}
  N_{\rm X} = \frac{{\rm ln}10\ \int A\ d\nu}{S^{\rm ref}_{\rm X}},
\end{equation}

\noindent where the ${\rm ln}10$ is needed to convert the integrated
absorbance, $A$, to optical depth and $S^{\rm ref}_{\rm X}$ is the
experimental RAIR band strength of species X. Transmission band
strengths available from the literature cannot be used, because the
total number of molecules probed with RAIRS cannot directly be related
to a value probed in a transmission absorption experiment.  This is
because the incident beam goes through the ice layer twice with an
angle to the surface. Instead, values of $S^{\rm ref}_{\rm X}$ have
been calculated from a calibration experiment without hydrogenation
where the deposition rate is 1.0$\times$10$^{-7}$~mbar~s$^{-1}$ to
grow a layer of typically 10 to 20~ML and where the sticking
probability is assumed to be 1. It should be noted that these
experimental values can differ between different experimental set-ups
and even for the same set-up over time, because they are determined by
intrinsic properties such as the alignment of the system. They should
therefore not be used to compare directly with observations of
interstellar ices and even a comparison between different experiments
should be made with caution. The values for $S^{\rm ref}_{\rm X}$ as
well as the spectral assignments of the vibrational modes are
summarized for all species in Table~\ref{int}.

There are several contributions to the uncertainty in the
experimentally measured band strengths. The largest fraction comes
from the actual deposition and sticking onto the surface. Since our
experiment does not contain a micro-balance, the absolute number of
molecules on the surface is not known and the possibility that some
molecules freeze out on other surfaces than the Au substrate cannot be
ruled out. This leads to systematic errors for $S_{\rm X}^{\rm ref}$
but there is no effect on the relative error for $S_{\rm X}^{\rm ref}$
between different experiments with the same ice morphology. Based on
experiments aimed at the same ice thickness, inaccuracies in the
deposition flow are estimated to be $\sim$30\%. Since the integrated
area of the absorption band is determined very precisely for the
deposited species, the column densities can be accurately normalized,
but the derived band strengths cannot. Another source of uncertainty
for ice mixtures is the precision of the mixing ratio, which is of the
order of 10\% \citep[see also][]{oberg2007,bisschop2007b}. In summary,
the actual uncertainty on the band strengths is substantial and ranges
from 30--40\%, but the relative uncertainty for ices with the same
morphology is less than 5\%.

\begin{table}
  \caption{Overview of the integrated frequency ranges with corresponding spectral assignment and band
    strength. The error on $S^{\rm ref}_{\rm X}$ amounts to $\sim$30-40\%.}\label{int}
\begin{center}
\begin{tabular}{llll}
  \hline
  \hline
  Species & Mode & Integration range & $S^{\rm ref}_{\rm X}$\\
  &             & (cm$^{-1}$)         & (cm~molecule$^{-1}$)\\
  \hline
  $^{13}$C$^{18}$O$_2$$^a$ &  $\nu_3$ & 2265--2245 & 1.3(-17)\\
  $^{13}$C$^{18}$O$_2$$^b$ &          & 2265--2245 & 8.8(-17)\\
  $^{13}$C$^{18}$O$_2$$^c$ &          & 2265--2245 & 4.8(-17)\\
  $^{13}$C$^{18}$O$_2$$^d$ &          & 2265--2245 & 4.3(-17)\\
  $^{13}$C$^{18}$O$_2$$^e$ &          & 2265--2245 & 3.2(-17)\\
  $^{13}$C$^{18}$O$_2$$^f$ &          & 2265--2245 & 2.7(-17)\\
  CO$^e$ & $\nu_1$ & 2160--2120 & 9.5(-18)\\
  CO$^f$ &         & 2160-2120  & 6.4(-18)\\
  HCOOH &  $\nu_3$ & 1800--1550 & 1.3(-16)\\
  CH$_3$CHO & $\nu_7$ & 1745--1719 & 8.0(-18)\\
  CH$_3$CHO &         & 1365--1330 & 2.8(-18)\\
  CH$_3$OH  & $\nu_8$ & 1060--990 & 1.1(-17)\\
  CH$_4$    & $\nu_4$ & 1320--1290 & 8.2(-18)\\
  \hline
\end{tabular}
\end{center} 

$^a$pure $^{13}$C$^{18}$O$_2$, $^b$22:78\%
$^{13}$C$^{18}$O$_2$:H$_2$O, $^c$39:61\% $^{13}$C$^{18}$O$_2$:H$_2$O,
$^d$48:52\% $^{13}$C$^{18}$O$_2$:H$_2$O, $^e$45:55\%
$^{13}$C$^{18}$O$_2$:CO, $^f$80:20\% $^{13}$C$^{18}$O$_2$:CO.
\end{table}

\subsection{Reaction rate calculations}
\label{rate_sec}

The method for calculating reaction rates is described in detail by
\citet{fuchs2007}. In short, a species X can react with H-atoms to
form species Z through: X + H $_{\longrightarrow}^{\ k_0}$ Z. The
column density of X that has reacted, $N_{\rm X}(t)$, is given by:

\begin{equation}
\label{eq1}
\frac{dN_{\rm X}(t)}{dt} = -k_0 N_{\rm H} N_{\rm X},
\end{equation}

\noindent where $k_0$ stands for the reaction constant and $N_{\rm H}$
for the surface density of H-atoms. In our experiment the H-atom flux
is kept constant and corresponds to 7.8$\times$10$^{13}$ or
5.0$\times$10$^{14}$~cm$^{-2}$~s$^{-1}$. Furthermore, the atoms have a
certain penetration depth as observed in experiments of H-atom
reactions with CO \citep{watanabe2004,fuchs2007} which means that not
all of the deposited parent species are available for the
reaction. Consequently, $N_{\rm X}(t)$ is calculated via:

\begin{equation}\label{eq2}
N_{\rm X}(t) = N_{\rm X}(0)\ \alpha_0\ (1-e^{-\beta_0 t}),
\end{equation}

\noindent where $\alpha_0$ is the fraction of $N_{\rm X}(0)$ that is
available to react and $\beta_0$ (in min$^{-1}$) corresponds to $k_0
N_{\rm H} /60$ (the factor 60 comes from the conversion of seconds to
minutes). In specific cases only upper limits on reaction rates are
calculated and then it is assumed that $\alpha_0$ only includes the
uppermost ice layer. Limits based on the column density decrease after
1~minute give the most conservative upper limit on $\beta_0$. Similar
to the cases described by \citet{fuchs2007}, fits to the reaction rate
differ when they are made over the complete time-period of the
experiment (hrs) or only over a shorter period (minutes).
 
\subsection{TPD analysis and calculation of the production yield}\label{yieldsec}

The TPD data provide complementary information on the reaction yields
and are important, in particular, for those molecules that are not
accurately determined by RAIRS. The TPD data are fitted by second
order polynomial baselines. The temperature range over which a
baseline is fitted depends on the desorption temperature of a specific
species. Calibration experiments of pure ices with a known number of
molecules have been performed by measuring the corresponding
integrated area of the mass spectrometer signal. The number of
molecules $N_{\rm Z}$ in other experiments has been determined by
comparison of the integrated signal to the number of molecules in the
calibration experiments. Since pure ices are needed for the
calibration, no accurate yields can be calculated for H$_2$CO which is
not readily available as a pure ice due to polymerization (see
\S~\ref{ch3cho_res}). The yield, $Y_{\rm Z}$, of the newly formed
species Z in \% can then be calculated through:

\begin{equation}\label{yieldeq}
Y_{\rm Z} = \frac{N_{\rm Z}}{\alpha_0\ N_{\rm X}(0)},
\end{equation}

\noindent where $N_{\rm Z}$ is the column density as derived from the
TPD data, $\alpha_0$ is taken from the fit of the RAIRS data and
$N_{\rm X}(0)$ is the initial column density of the precursor species
as derived from the RAIR spectra.

\section{CO$_2$ containing ices}
\label{co2_sec}
\subsection{Results}\label{co2_res}

\begin{figure}\centering
\includegraphics[width=8cm]{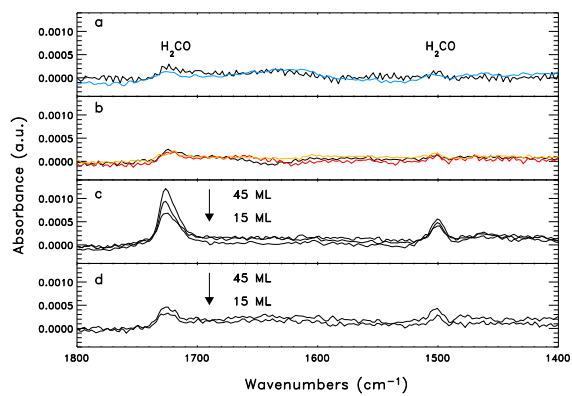}
\caption{Difference spectra of the 1800--1400~cm$^{-1}$ range after
  180~minutes of H-atom bombardment. (a) 15~ML pure C$^{18}$O$_2$ ice
  (black line) and $^{13}$C$^{18}$O$_2$ (grey line), (b) 15~ML
  CO$_2$:H$_2$O 39:61\% (black line), 22:78\% (grey line) and 48:52\%
  (light grey), (c) 45:55\% CO$_2$:CO mixtures for 15, 30 and 45~ML
  ice thickness and (d) 80:20\% CO$_2$:CO mixture for 15 and
  30~ML. The temperature of the ice is $\sim$14.5~K. The arrows
  indicate how the absorbance of the H$_2$CO bands decreases with
  decreasing thickness.}\label{spec}
\end{figure}

Pure ices of CO$_2$ are bombarded with H-atoms to search for the
possible formation of HCOOH following the reaction route as shown in
Fig.~\ref{reaction-scheme}. The reactivity of CO$_2$ can be tested by
recording a decrease in the CO$_2$ RAIR signal and by monitoring any
reaction products. Unfortunately, the CO$_2$ 2300~cm$^{-1}$ ice band
is difficult to quantify, because this band overlaps with
rotation-vibration transitions of CO$_2$ present in the purge gas in
our spectrometer. The use of $^{13}$C$^{18}$O$_2$ isotopic species
does not improve this situation. Therefore we have focused on the
strongest HCOOH band, the C=O stretching mode, at 1710~cm$^{-1}$, to
monitor formic acid formation. The observed difference RAIR spectrum
of $^{13}$C$^{18}$O$_2$ bombarded by H-atoms has been compared to that
of $^{12}$C$^{18}$O$_2$ in Fig.~\ref{spec}a. Both spectra show very
weak features at 1730~cm$^{-1}$ and 1500~cm$^{-1}$, but none at
positions typical for HCOOH. The detected bands, however, occur at
exactly the same positions as previously seen for H$_2$CO when formed
upon CO hydrogenation \citep[e.g.,][]{watanabe2004}. Furthermore, the
features do not shift when a different isotopic species is used, which
is expected when the formation involves CO$_2$ ice. In the TPD spectra
of CO$_2$ ices bombarded with H-atoms for 3~hrs (not shown) mass 29
and 30~amu desorb in two steps around 100~K and 140~K, and the
formation of other species is not detected through TPD. These
temperatures and masses are identical to what was observed by
\citet{fuchs2007} for H$_2$CO desorption from CO ices bombarded by
H-atoms. During H-atom bombardment an increase in the mass 28~amu
signal is observed, which is likely due to degassing of both CO and
N$_2$ from the metal parts of our experiment. It is therefore
plausible that the measured low level of H$_2$CO formation observed in
the experiment originates from hydrogenation of background gaseous CO
and is not related to the CO$_2$ ice.

To test whether the presence of H$_2$O affects the reactivity of
CO$_2$ ice upon H-atom bombardment as has been observed for CO in
CO:H$_2$O mixtures \citep{watanabe2004,fuchs2007}, mixtures of 22:78\%
to 48:52\% CO$_2$:H$_2$O have been investigated (see
Fig.~\ref{spec}b). Like in the experiments with pure CO$_2$, weak RAIR
features of similar intensity are observed at 1730 and
1500~cm$^{-1}$. Again we assign these features to the C=O stretching
mode and the C-H bending mode of H$_2$CO. Thus, as for the pure ices,
a small amount of background CO accretes and forms H$_2$CO. Within the
sensitivity of our experiment, we conclude that CO$_2$ does not react
with H-atoms even if mixed with H$_2$O.

Finally, mixtures of $^{12}$C$^{16}$O and $^{13}$C$^{18}$O$_2$ are
studied to determine which species is more likely to react upon H-atom
bombardment: CO or CO$_2$. Since CO hydrogenation reactions were
previously reported in the literature \citep{watanabe2004,fuchs2007},
the answer to this question must be CO. In Fig.~\ref{spec}c and d, the
resulting difference spectra for CO$_2$:CO mixtures are shown for
different ice thicknesses and with mixture concentrations of 45:55\%
CO$_2$:CO and 80:20\% CO$_2$:CO, respectively. Similar to H-atom
bombardment of pure CO$_2$ ices and CO$_2$:H$_2$O mixtures, no
evidence for HCOOH formation is observed in CO$_2$:CO mixtures. In
contrast, CO does react with H-atoms to form H$_2$CO and CH$_3$OH as
is evidenced by the presence of strong H$_2$CO absorption features at
1730 and 1500~cm$^{-1}$ and CH$_3$OH at 1030~cm$^{-1}$. This is
consistent with H-atom bombardment experiments for pure CO and
CO:H$_2$O mixtures by \citet{watanabe2004} and \citet{fuchs2007}. The
complementary TPD data show the same picture of no HCOOH formation and
clear H$_2$CO and CH$_3$OH formation from CO. Other products than the
precursor and product species H$_2$CO and CH$_3$OH are not observed.

\subsection{Reaction rates}
\label{rate_co2}

\begin{table}
  \caption{Upper limits on the reaction/destruction rates for HCOOH and
    CO$_2$. The uncertainty on $k_0$ amounts to a factor 2.}\label{dest}
\begin{center}
\begin{tabular}{llll}
  \hline 
  \hline
  Species & Ice matrix & T$_{\rm ice}$ & $k_0$ \\
  &          & (K)  & (cm$^2$~s$^{-1}$)\\
  \hline
  CO$_2$  & pure       & 14.5 & $\leq$6.2(-17)\\
  CO$_2$  & $^{13}$C$^{18}$O$_2$:H$_2$O 39:61\% & 14.5 & $\leq$6.0(-17)\\
  CO$_2$$^a$  & $^{13}$C$^{18}$O$_2$:H$_2$O 39:61\% & 14.5 & $\leq$3.8(-17)\\
  CO$_2$  & $^{13}$C$^{18}$O$_2$:H$_2$O 22:78\% & 14.5 & $\leq$6.7(-17)\\
  CO$_2$  & $^{13}$C$^{18}$O$_2$:H$_2$O 48:52\% & 14.5 & $\leq$3.2(-17)\\
  CO$_2$  & $^{13}$C$^{18}$O$_2$:H$_2$O 39:61\% & 14.5 & $\leq$6.0(-17)\\
  \hline
  HCOOH   & pure       & 12.5 & $\leq$2.3(-17)\\
  \hline
\end{tabular}
\end{center}

$^a$ Limit derived for experiment without bombardment.
\end{table}

No hydrogenation products of CO$_2$, specifically HCOOH, are observed
within the experimental sensitivity. The limit on the formation
reaction rate for HCOOH from CO$_2$ is
$\leq$7.0$\times$10$^{-17}$~cm$^{2}$~s$^{-1}$ based on the limit on
the column density for HCOOH after 1~min of H-atom bombardment for all
ice morphologies (see \S~\ref{rate_sec} for the derivation and
Table~\ref{dest} for the individual values for each experiment).

\begin{figure}\centering
\includegraphics[width=8cm]{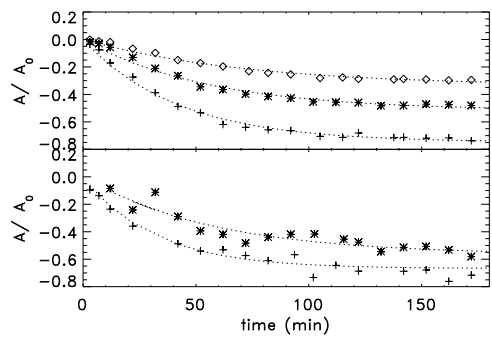}
\caption{$A$/$A_0$ for the CO 2140~cm$^{-1}$ band in CO$_2$:CO
  mixtures at 14.5~K for the 45:55\% (upper panel) and 80:20\% (lower
  panel). The symbols refer to 15~ML $+$, 30~ML $\ast$, and 45~L
  $\Diamond$.  The dotted lines indicate the fits to the
  data.}\label{cofit}
\end{figure}

In Figure~\ref{cofit} the absorbance divided by the initial absorbance
at $t=$0, $A/A_0$, is shown for the 45:55\% and 80:20\% CO$_2$:CO ice
mixtures. The data are fitted as described in \S~\ref{rate_sec} and
the fits are indicated in Fig.~\ref{cofit} with dotted lines. The
resulting values for $\alpha_0$ and $\beta_0$ as well as the $k_0$ are
given in Table~\ref{corate}. Since the H$_2$CO band strength could not
be determined accurately in these experiments only $\alpha_0$ and
$\beta_0$ are fitted. Clearly $\alpha_0$, i.e., the fraction of CO
molecules available for reaction, decreases with increasing ice
thickness.

\begin{table}
\centering
\caption{Values for $\alpha_0$, $\beta_0$, and the reaction rate $k_0$
  for CO in CO:CO$_2$ mixtures at ice temperatures of 14.5~K upon H-atom bombardment. The uncertainties for $\alpha_0$ and $\beta_0$ amount to 10--20\% and for $k_0$ are a factor 2.}\label{corate}
\begin{tabular}{lllll}
\hline
\hline
Ice mixture & Ice thickness & $\alpha_0$ & $\beta_0$ & $k_0$\\
            & CO$_2$/total  &            &           & \\
            & (ML)          &            & (min$^{-1}$) & (cm$^2$~s$^{-1}$)\\
\hline
45:55\% CO$_2$:CO & 8/15       & 0.75       & 0.024        & 2.9(-15)\\
45:55\% CO$_2$:CO & 17/30      & 0.51       & 0.018        & 2.2(-15)\\
45:55\% CO$_2$:CO & 25/45      & 0.34       & 0.014        & 1.7(-15)\\
\hline
80:20\% CO$_2$:CO & 3/15       & 0.67       & 0.032        & 3.8(-15)\\
80:20\% CO$_2$:CO & 6/30       & 0.57       & 0.017        & 2.0(-15)\\
\hline
\end{tabular}
\end{table}

\subsection{Discussion and conclusion}\label{co2_disc}

\begin{figure*}\centering
\includegraphics[width=12cm]{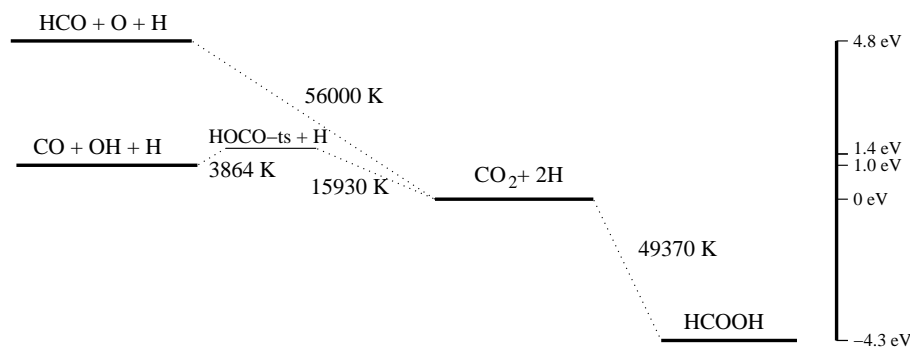}
\caption{Potential energy scheme for CO$_2$ dissociation and
  hydrogenation. The relative energies are based on the heats of
  formation at 0~K. The scale in electronvolts is indicated on the
  right. The heats of formation are derived from \citet{cox1989} for
  H, CO$_2$, CO and O, from \citet{ruscic2002} for OH, from
  \citet{gurvich1989} for HCO and HCOOH and from \citet{lakin2003} for
  the HOCO transition state(ts).}\label{energy-co2}
\end{figure*}

Previously, CO$_2 +$H reactions have been studied by
\citet{milligan1971} in UV-photolysis experiments of Ar:CO$_2$:H$_2$O
matrices. Only the formation of more oxygen-rich species such as
CO$_3$ has been observed, but not HCOOH. Although these results cannot
be compared directly with ours (see \S~\ref{plab}), our findings are
consistent with theirs that CO$_2$ does not react readily with
H-atoms. To explain the lack of CO$_2$ hydrogenation reactions,
Fig.~\ref{energy-co2} presents the relative formation energies of
possible products. Reaction of CO$_2$ with H-atoms to either HCO$+$O
or CO$+$OH is energetically highly unfavorable. This is due to the
HOCO transition state being $\sim$15930~K (1.4~eV) higher in energy
compared to CO$_2$$+$H in the gas phase \citep{lakin2003}. On the
other hand, a hydrogenation reaction could be expected based on the
higher heat of formation of CO$_2$$+$2H with respect to HCOOH of
$\sim$49370~K (4.3~eV). Indeed, in the chemical physics literature
CO$_2$ is found to hydrogenate to HCOOH on ruthenium and iridium
catalysts through formate complexes with the catalyst
\citep{ogo2006}. The chemically bonded formate species subsequently
reacts with H$_3$O$^+$ to form HCOOH. This reaction mechanism requires
acidic species as well as a catalytic surface that are not present in
the current experiment. \citet{hwang2004} calculated a potential
energy surface for the gas phase H$_2$~$+$~CO$_2$ reaction. The HCOOH
end product is higher in energy than the initial species by
$\sim$2600~K (0.22~eV), but the CO$_2$~$+$~2H reaction is exothermic
(see Fig.~\ref{energy-co2}). For the reaction of CO$_2$ with H-atoms
or H$_2$ the same transition state H$_2$CO$_2$, a complex cyclic
structure, has to be overcome. This transition state lies
35000--37000~K (3.0--3.2~eV) above the starting point,
CO$_2$~$+$~H$_2$, but below CO$_2$~$+$~2~H. This step can therefore
not be rate-limiting for the CO$_2$$+$ 2H~$\rightarrow$~HCOOH
reaction. Since the reaction is clearly not observed in our
experiment, the rate determining step must be the addition of the
first hydrogen atom to CO$_2$ to form HCO$_2$ and the barrier to this
reaction must be too high to be overcome for the ice temperatures of
12--60~K as used in our experiments.

Several recent studies have focused on reactions of CO $+$ H leading
to the formation of H$_2$CO and CH$_3$OH. \citet{fuchs2007} find that
even for the lowest ice thicknesses of 1 to 2~ML 30\% of the ice is
hidden from the impinging H-atoms. At higher thicknesses $\alpha_0$
increases and there is a maximum layer thickness of 12~ML of CO ice
that can react with H-atoms. The behavior of $\alpha_0$ for our
CO$_2$:CO mixtures is consistent with this picture, although at
maximum 7$\pm$3~ML and 8$\pm$3~ML of the CO ice reacts for the 45:55\%
and 80:20\% CO$_2$:CO mixtures, respectively. In other words mixing CO
with CO$_2$ does not cause more CO to be ``hidden'' from the H-atom
exposure.

Our reaction rates, $\beta_0$, of e.g., 0.032~min$^{-1}$ in CO:CO$_2$
80:20\% with 15~ML total ice thickness at 14.5~K are similar to those
found for pure CO ices by \citet{fuchs2007} of $\sim$0.030~min$^{-1}$
($\sim$0.023~min$^{-1}$ when converted to our assumption for the
initial number molecules) for a similar amount of CO of 11~ML at
15~K. These rates are the same within the 30-40\%
uncertainty. However, for CO:H$_2$O 1:5 mixtures \citet{fuchs2007}
find a value of $\sim$0.11~min$^{-1}$ (0.083~min$^{-1}$) for 12~ML at
15~K indicating that the reaction rates for CO hydrogenation are
significantly higher in mixtures with H$_2$O ice. The similarity
between the reaction rate of CO in mixtures with CO$_2$ and pure CO
ices and the difference between those and CO:H$_2$O ice mixtures can
be explained by CO and CO$_2$ only interacting through weak
Van~der~Waals forces.  The electronic structure of the CO molecule
will therefore not differ significantly in mixtures with CO$_2$ from
pure CO ices. H$_2$O on the other hand has a stronger dipole moment of
1.85~D compared to zero and 0.11~D for CO$_2$ and CO, respectively,
and forms hydrogen bonds. Furthermore it is known that CO strongly
interacts with and influences the band strengths of H$_2$O molecule
\citep{bouwman2007}. Thus, the electronic structure of the CO molecule
will be perturbed in mixtures with H$_2$O, strongly affecting the
reaction rate of CO with H-atoms. In summary the presence of CO$_2$ in
ice mixtures with CO does not strongly affect the reactivity of CO
with H-atoms.

\section{HCOOH containing ices}
\label{hcooh_sec}

\subsection{Results}
\label{hcooh_res}

\begin{figure}\centering
\includegraphics[width=8cm]{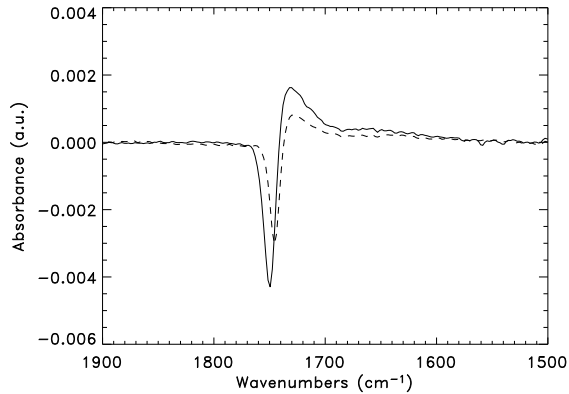}
\caption{The difference spectrum of the HCOOH $\nu_{\rm S}$(C=O)
  stretch for HCOOH bombarded for 4~hrs with H-atoms at 12~K (solid)
  and at 40~K (dashed). }\label{hcooh}
\end{figure}

Figure~\ref{hcooh} shows the difference spectrum for $\nu_{\rm
  S}$(C=O) at $\sim$1710~cm$^{-1}$ of pure HCOOH ice bombarded with
H-atoms as well as control experiments with bombardment of H$_2$
molecules at 12~K \citep[for an overview of all infrared features of
HCOOH see][]{cyriac2005}. The growth of an infrared feature around
$\sim$1050~cm$^{-1}$ indicative for CH$_2$(OH)$_2$ formation has not
been observed \citep[for an overview of the infrared features of
CH$_2$(OH)$_2$ see][]{lugez1994}. A decrease on the blue side of the
$\nu_{\rm S}$(C=O) mode at 1710~cm$^{-1}$ of HCOOH is seen at
1750~cm$^{-1}$ as well as an increase at 1730~cm$^{-1}$, which means
that the overall HCOOH band profile changes slightly. At
1730~cm$^{-1}$, the C=O stretch for H$_2$CO is located, but other
features of H$_2$CO, such as the 1500~cm$^{-1}$ band, are missing.
The decrease corresponds to $<$0.1~ML derived from our calculated RAIR
band strength. These features are present in difference spectra for
HCOOH ice bombarded with H-atoms at 12 and 40~K. A similar shift is
seen for transmission infrared experiments with pure HCOOH ice that is
heated to $\sim$60~K \citep{bisschop2007b}. At the same time the
$\nu_{\rm S}$(CH) and $\nu_{\rm S}$(OH) vibrational modes increase due
to conversion of HCOOH in dimeric form to HCOOH organized in
chains. In the RAIRS spectra these bands are also seen to
increase. Furthermore, the same change in RAIR profile is found for
HCOOH ices of 40~K, where H-atoms cannot stick any longer onto the
surface, but can only collide. Since the ice has a temperature of
40~K, the reorganization of the ice is less and consequently the
signal of the difference spectrum is smaller. In conclusion, the RAIR
data do suggest that some restructuring takes place in the surface but
no reaction.

With TPD the masses of 48 (CH$_2$(OH)$_2$), 46 (HCOOH), 45 (HCOO), 44
(CO$_2$), 32/31 (CH$_3$OH), 30/29 (H$_2$CO), and 28~amu (CO) have been
monitored during warm-up. No products are detected at 48, 32, 31, or
30~amu to upper limits of $<$0.01~ML, indicating that HCOOH is neither
hydrogenated nor dissociated. Thus, consistent with the lack of a
1500~cm$^{-1}$ H$_2$CO absorption feature in the RAIRS data, no
evidence for H$_2$CO formation is observed in the TPD experiment.  The
detected masses 45, 44, and 29~amu are assigned to HCOOH dissociating
in the mass spectrometer, because the same relative mass ratios are
seen for a TPD spectrum of pure HCOOH ice that is not bombarded by
H-atoms. We conclude that within the limits of our experimental set-up
the reaction of HCOOH with H-atoms is not efficient at 12~K.

\subsection{Reaction rates}\label{hcooh_rate}

Since no unambiguous evidence for HCOOH destruction in the ice is
found, it is only possible to derive an upper limit on its reaction
rate, presented in Table~\ref{dest}. It is clear that the HCOOH
destruction rates are below
2.3$\times$10$^{-17}$~cm$^2$~s$^{-1}$ as derived from the limit
on the column density after 1~min of H-atom bombardment (see
\S~\ref{rate_sec}). As for CO$_2$ these reaction rates are very
low.

\subsection{Discussion and conclusion}\label{hcooh_disc}
\begin{figure}\centering
\includegraphics[width=8cm]{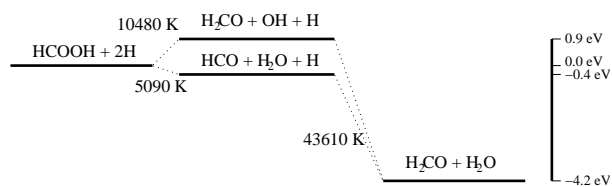}
\caption{Potential energy scheme for HCOOH hydrogenation. The energies
  are based on the heats of formation at 0~K. The energy scale in
  electronvolts is indicated on the right. The heats of formation are
  derived from \citet{gurvich1989} for HCO, H$_2$CO and HCOOH, from
  \citet{ruscic2002} for OH and from \citet{cox1989} for H and
  H$_2$O.}\label{energy-hcooh}
\end{figure}

In chemical physics literature HCOOH hydrogenation on catalytic
surfaces has been shown to lead to decomposition of HCOOH rather than
methanediol formation \citep{benitez1993}. HCOOH adsorbs onto such a
surface as HCOO$^-$ and H$^+$ which can be further hydrogenated. The
catalytic surface, however, clearly affects the end products and
overcomes a reaction barrier that prohibits spontaneous
decomposition. If hydrogen atom addition and dissociation occur
simultaneously, C--O bond cleavage is more energetically favorable (as
shown in Fig.~\ref{energy-hcooh}). However, it is clear from the
results in \S~\ref{hcooh_res} that no H$_2$O and H$_2$CO formation
occurs. Thus a high barrier for H-addition to HCOOH must exist for
both mechanisms and HCOOH $+$ H reactions in the ice are inefficient.

\section{CH$_3$CHO containing ices}
\label{ch3cho_sec}

\begin{figure}\centering
\includegraphics[width=8cm]{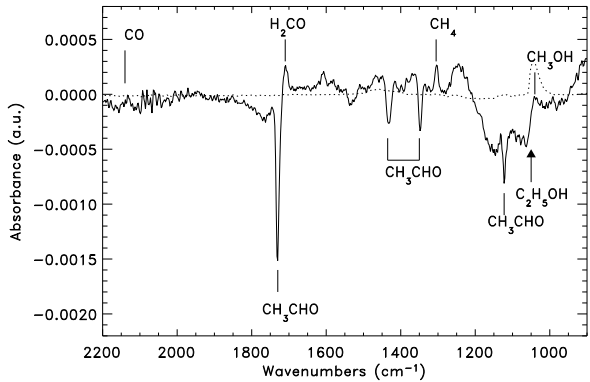}
\caption{The difference spectrum of CH$_3$CHO after 3~hrs of
  bombardment with H-atoms. The negative peaks correspond to CH$_3$CHO
  destruction, the positive wing around 1700~cm$^{-1}$ is assigned to
  the $\nu_{\rm S}$(C=O) mode of H$_2$CO and the features at
  1300~cm$^{-1}$ and 1030~cm$^{-1}$ to $\nu_{\rm D}$(CH$_4$) of CH$_4$
  and $\nu_{\rm S}$(C--O) of CH$_3$OH, respectively. The arrow
  indicates the position where the strongest C$_2$H$_5$OH absorption
  is expected. The dotted line indicates the RAIR spectrum for pure
  CH$_3$OH, showing that the feature detected at 1030~cm$^{-1}$
  matches that of pure CH$_3$OH.}\label{ch3chobomb}
\end{figure}

\subsection{Results}
\label{ch3cho_res}

The infrared spectroscopic features detected for pure CH$_3$CHO ice
match with those detected by \citet{bennett2005a} and
\citet{moore1998,moore2003}. The strongest CH$_3$CHO band is the C=O
stretching mode, $\nu_{\rm S}$(C=O), at 1728~cm$^{-1}$
(5.79~$\mu$m). During H-atom bombardment the intensity decreases, but
a small positive wing is observed at 1710~cm$^{-1}$ (see
Fig.~\ref{ch3chobomb}). This band is assigned to the C=O stretching
mode of H$_2$CO. Other CH$_3$CHO features (e.g., the umbrella
deformation mode, $\nu_{\rm D}$, at 1345~cm$^{-1}$) also decrease and
new bands appear at 1030 and 1300~cm$^{-1}$ that are attributed to the
C--O stretching mode of CH$_3$OH and the deformation mode of CH$_4$,
respectively. No clear absorption is observed at 1050~cm$^{-1}$, where
the strongest C$_2$H$_5$OH band, the C--O stretching mode, is
expected. Since this frequency region is particularly problematic in
our detector, the detection upper limit on $N$(C$_2$H$_5$OH) amounts
to only 3$\times$10$^{15}$~molecules~cm$^{-2}$, i.e., 3~ML. Another
strong band of C$_2$H$_5$OH is expected at 3.5~$\mu$m. Unfortunately,
this feature overlaps with a number of CH$_3$OH modes. Broad weak
features are indeed detected in this range, but due to the complexity
of both C$_2$H$_5$OH and CH$_3$OH absorptions and the relatively weak
signal this cannot be used to determine whether C$_2$H$_5$OH is
present. Additionally, it is important to note that no strong features
are observed around 2140~cm$^{-1}$, where both CO and CH$_2$CO have
infrared features. This is perhaps not surprising, because the
formation of CO would involve not only the breaking of a C--C bond,
but also hydrogen-abstraction, which is not very likely in this
hydrogen-rich environment. The formation of ketene, CH$_2$CO, is even
less likely because its formation is strongly endothermic.

\begin{figure}\centering
\includegraphics[width=8cm]{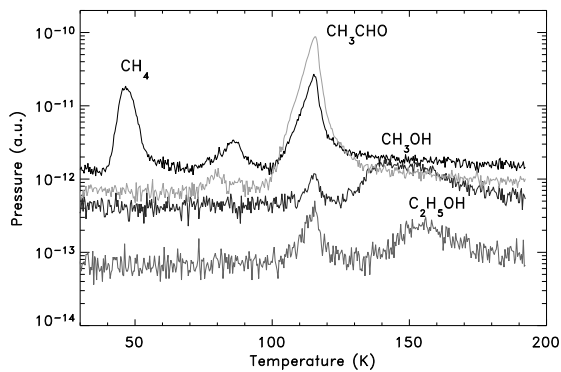}
\caption{The TPD spectrum for 40~ML of CH$_3$CHO bombarded with
  H-atoms for 3~hrs at 14.5~K. The black line refers to the 16~amu
  signal (CH$_4$), dark grey line to 31~amu (CH$_3$OH), grey to 46~amu
  (C$_2$H$_5$OH) and light grey to 44~amu (CH$_3$CHO).}\label{tpd}
\end{figure}

The formation of CH$_4$, H$_2$CO and CH$_3$OH is corroborated by the
TPD spectra, where 16, 30 and 31 amu mass peaks at 45~K, 100~K and
140~K are found, respectively (see Fig.~\ref{tpd} for CH$_4$ and
CH$_3$OH). The peaks for 16~amu at higher temperatures are due to
O-atoms detected by the mass spectrometer when other molecules
dissociate. The desorption temperatures for 16 and 31~amu are similar
to those of pure CH$_4$ and CH$_3$OH ice confirms their RAIR
detection. The TPD spectra and desorption temperatures of 29~amu are
consistent with the desorption temperatures for H$_2$CO found by
\citet{watanabe2004}. In addition, a TPD desorption peak is located at
$\sim$160~K for masses 45 and 46~amu (see Fig.~\ref{tpd}). This is
assigned to C$_2$H$_5$OH desorption based on a comparison with the TPD
of pure non-bombarded C$_2$H$_5$OH ices. In summary, a fraction of
CH$_3$CHO, below the infrared detection limit of the 1050~cm$^{-1}$
band, is converted to C$_2$H$_5$OH and a larger fraction forms CH$_4$,
H$_2$CO and CH$_3$OH. So even though the conversion of acetaldehyde to
ethanol is not complete, it is important to note that a pathway in the
proposed hydrogenation scheme by \citet{tielens1997} is experimentally
confirmed.

\subsection{Reaction rates and production yields}\label{sec_rate_ch3cho}

\begin{figure}\centering
\includegraphics[width=8cm]{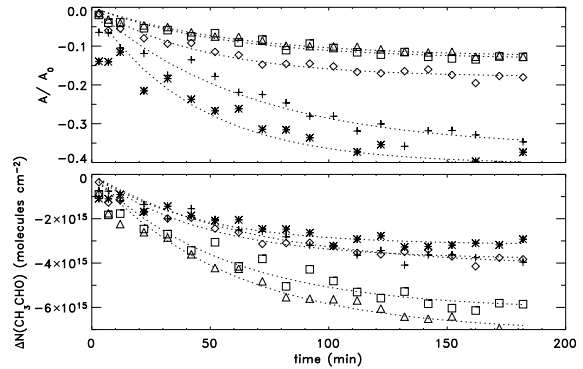}
\caption{The $A$/A$_0$ ratio (upper panel) and $\Delta N$(CH$_3$CHO)
  (lower panel) for the CH$_3$CHO 1345~cm$^{-1}$ band for different
  ice thicknesses and a constant ice temperature of 14.5~K. The
  symbols refer to 11.4~ML ($+$), 7.8~L ($\ast$), 21.2~L ($\Diamond$),
  45.8~ML ($\Box$) and 56.0~ML ($\bigtriangleup$). The dotted lines
  indicate the fits to the data.}\label{ch3chofit}
\end{figure}

\begin{figure}\centering
\includegraphics[width=8cm]{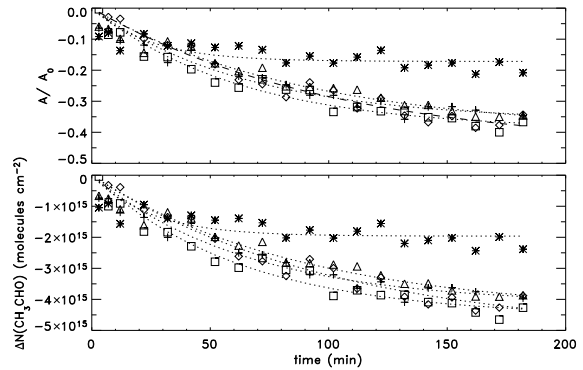}
\caption{The $A$/A$_0$ ratio (upper panel) and $\Delta N$(CH$_3$CHO)
  (lower panel) for the CH$_3$CHO 1345~cm$^{-1}$ band for different
  ice temperatures and a constant ice thickness of 11.3~ML. The
  symbols refer to 14.5~K ($+$), 12.4~K ($\ast$), 15.8~K ($\Box$),
  17.4~K ($\Diamond$) and 19.3~K ($\bigtriangleup$). The dotted lines
  indicate the fits to the data.}\label{ch3chofit2}
\end{figure}

The value for $N$(CH$_3$CHO) as derived from the $\nu_{\rm
  D}$(umbrella) spectral feature at 1345~cm$^{-1}$ is shown in
Fig.~\ref{ch3chofit} as a function of time for different ice
thicknesses at 14.5~K. Also shown are the fits to the data. The
$\nu_{\rm D}$(umbrella) mode is chosen for analysis rather than the
1728~cm$^{-1}$ band, because the latter overlaps with the $\nu_{\rm
  S}$(C=O) of H$_2$CO at 1720~cm$^{-1}$. Clearly, the absolute amount
of CH$_3$CHO that can react increases with ice thickness, whereas
$A/A_0$ decreases. The temperature behavior is more complex and is
shown in Fig.~\ref{ch3chofit2}.

\begin{figure}\centering
\includegraphics[width=8cm]{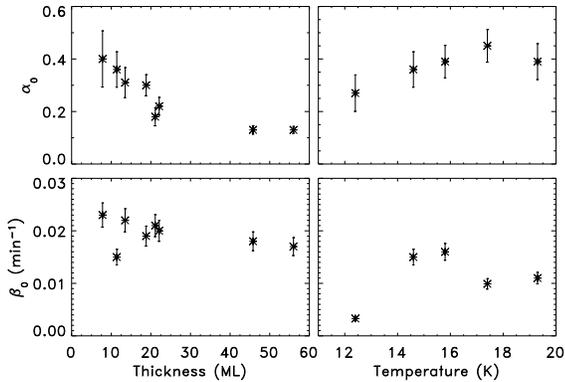}
\caption{The $\alpha_0$ and $\beta_0$ dependencies of the
  CH$_3$CHO$+$H reaction on thickness and temperature. The temperature
  of the ices for the thickness dependence experiments is constant at
  14.5~K. The ice thickness is similar for the temperature dependence
  experiments at 11--12~ML.}\label{params}
\end{figure}

The $\alpha_0$ and $\beta_0$ values derived from the fits as function
of the thickness and temperature are shown in Fig.~\ref{params}. The
values for $\alpha_0$ decrease with increasing thickness, but do not
depend on ice temperature within the measured regime. The latter is
not surprising as the CH$_3$CHO ice structure does not change between
15 and 75~K. The value for $\beta_0$ is independent of ice thickness,
but does depend on ice temperature. It is largest for ice temperatures
between 15--16~K, similar to the case of CO \citep{fuchs2007}. This is
expected as the maximum reactivity is mostly determined by the
mobility of H-atoms at the surface. At low temperatures H-atoms move
more slowly resulting in a lower reaction rate. At higher temperatures
the diffusion rate is higher but has to compete with an increased
evaporation rate.

\begin{table*}\centering
  \caption{Values for $\alpha_0$, $\beta_0$, the reaction rate $k_0$
    for CH$_3$CHO, and production yields, $Y$(X) upon H-atom bombardment. The uncertainties for $\alpha_0$ and $\beta_0$ amount to 10-20\%, for $k_0$ are a factor 2 and $Y$ 20\%.}\label{ch3chorate}
\begin{tabular}{lllll|lll}
  \hline
  \hline
  Ice temperature & Ice thickness & $\alpha_0$ & $\beta_0$ & $k_0$ & $Y$(C$_2$H$_5$OH) & $Y$(CH$_4$) & $Y$(CH$_3$OH)\\
  (K)             & (ML)          &            & (min$^{-1}$) & (cm$^2$~s$^{-1}$) & (\%) & (\%) & (\%)\\
  \hline
  14.5            & 7.8            & 0.40       & 2.3(-2)        & 2.8(-15) & 15 & 21\\
  14.5            & 11.4           & 0.36       & 1.5(-3)        & 1.8(-16) & 10 & 19\\
  14.6            & 13.5           & 0.31       & 2.2(-2)        & 2.6(-15) & 14 & 22 & 48\\
  14.5            & 18.8           & 0.30       & 1.9(-2)        & 2.3(-15) & 13 & 18\\
  14.5            & 21.2           & 0.18       & 2.1(-2)        & 2.5(-15) & 17 & 24\\
  14.6            & 22.1           & 0.22       & 2.0(-2)        & 2.4(-15) & 20 & 23 & 39\\
  14.6            & 45.8           & 0.13       & 1.8(-2)        & 2.2(-15) & 20 & 21 & 38\\
  14.6            & 56.0           & 0.13       & 1.7(-2)        & 2.0(-15) & 20 & 17 & 35\\
  15.8            & 11.6           & 0.39       & 1.6(-3)        & 1.9(-16) & 21 & 22 & 33\\
  17.4            & 11.3           & 0.45       & 9.9(-3)        & 1.2(-15) & 16 & 19\\
  19.3            & 11.2           & 0.39       & 1.1(-2)        & 1.3(-15) & 14 & 13 & 15\\
  \hline
\end{tabular}
\end{table*}

For the C$_2$H$_5$OH formation, only yields can be calculated from the
TPD data because the RAIR feature at 1050~cm$^{-1}$ overlaps with the
$\nu_{\rm S}$(CO) band of CH$_3$OH. The yields for CH$_4$, CH$_3$OH
and C$_2$H$_5$OH are given in Table~\ref{ch3chorate}. Even when
considering that there is a general quantitative uncertainty of
$\sim$10\%, it is clear that the summed yield of the different
products is not 100\%. This is most likely due to missing H$_2$CO
yields, because these are not reliably calibrated. Furthermore
$Y$(CH$_4$) is expected to be equal to $Y$(H$_2$CO$+$CH$_3$OH),
because CH$_3$OH is formed from H$_2$CO after CH$_3$CHO
dissociation. However, the CH$_3$OH yield is significantly higher than
CH$_4$ (see \S~\ref{discon}). The solid state C$_2$H$_5$OH yields
are $\leq$20\%.

\subsection{Discussion and conclusion}
\label{discon}
\begin{figure*}\centering
\includegraphics[width=16cm]{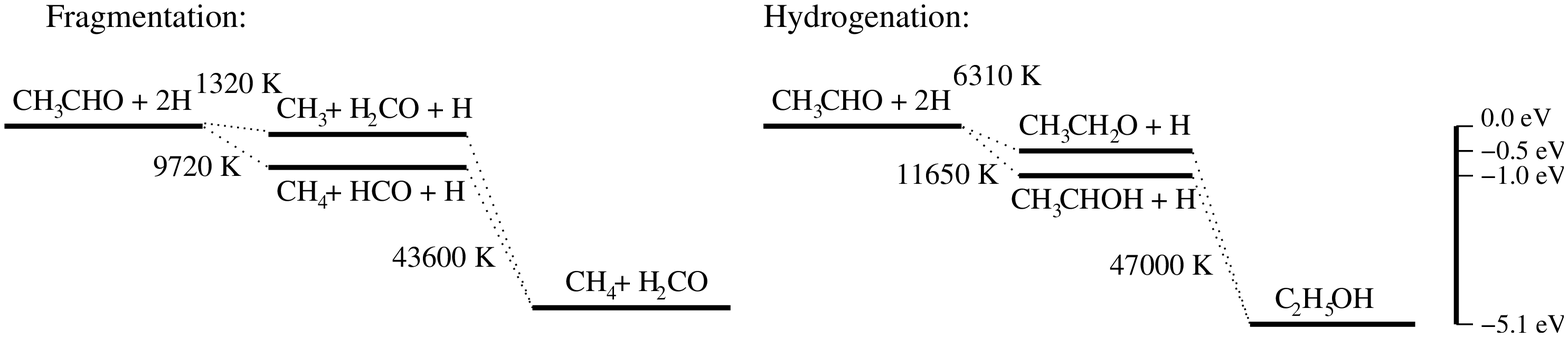}
\caption{Potential energy scheme for CH$_3$CHO fractionation and
  hydrogenation. The relative energies are based on the heats of
  formations at 0~K. An approximate energy scale in electronvolts is
  given on the right. The heats of formation are derived from
  \citet{wiberg1991} for CH$_3$CHO, \citet{cox1989} for H,
  \citet{gurvich1989} for CH$_3$, H$_2$CO, CH$_4$ and HCO,
  \citet{frenkel1994} for C$_2$H$_5$OH and \citet{matus2007} for the
  CH$_3$CHOH and CH$_3$CH$_2$O radicals.}\label{energy}
\end{figure*}

Previously, C$_2$H$_5$OH and CH$_3$CHO were shown to form in
interstellar ice analogues by photolysis of C$_2$H$_2$:H$_2$O mixtures
\citep{moore2001,wu2002}. Since in such experiments both OH and H
fragments are present with excess energy, it is difficult to
disentangle potential pure hydrogenation reactions and reactions
involving OH radicals. Indeed, \citet{moore2005} explain formation of
C$_2$H$_5$OH and CH$_3$CHO by reactions of C$_2$H$_5$ and C$_2$H$_3$
with OH, respectively. In this paper we focus on the reactions with
thermal H-atoms only.

Since H$_2$CO is known to react with H-atoms to CH$_3$OH
\citep{watanabe2004,hidaka2004} it is thus likely that the next more
complex aldehyde, acetaldehyde (CH$_3$CHO), will form ethanol
(C$_2$H$_5$OH). In Figure~\ref{energy} the relative heats of formation
at 0~K are shown
\citep{wiberg1991,cox1989,gurvich1989,frenkel1994,matus2007}. The
exothermicity of CH$_3$CHOH formation is higher than that of
CH$_3$CH$_2$O, but which of the species is more likely formed depends
on the reaction barriers. Subsequent formation of C$_2$H$_5$OH is
likely fast, because reactions of radicals with H-atoms commonly have
no activation barriers. As described in \S~\ref{sec_rate_ch3cho} only
a fraction of CH$_3$CHO is converted to C$_2$H$_5$OH, and a larger
fraction leads to CH$_4$, H$_2$CO, and CH$_3$OH formation. For
hydrogenation of CH$_3$CHO a C=O bond is converted to a C--O bond
instead of breaking a C--C bond. Since the C=O bond is intrinsically
stronger it is likely that the entrance channel to hydrogenation is
higher in energy compared to dissociation.

Thus H-atoms can break the C--C bond as well as the C=O bond to form
CH$_4$, H$_2$CO and CH$_3$OH or C$_2$H$_5$OH in ices as prepared
here. As shown in Fig.~\ref{energy} the formation of CH$_4$$+$HCO is
more exothermic than that for CH$_3$$+$H$_2$CO. Furthermore, the
energy released in this step is higher than the binding energy of
CH$_4$ to the surface, which is $\sim$700~K (0.06~eV). This likely
explains why the $Y$(CH$_4$) is lower than $Y$(CH$_3$OH$+$H$_2$CO),
because the formation energy is sufficient for CH$_4$ desorption. The
energy released during the formation of H$_2$CO, CH$_3$OH and
C$_2$H$_5$OH is even higher and may also cause a fraction of the
molecules to desorb.

\section{Astrophysical implications}
\label{disc}

Our experiments show that CO$_2$ reaction rates with H-atoms are very
low, making it an implausible route for HCOOH formation. A number of
other HCOOH formation routes are possible \citep[see
e.g.,][]{milligan1971,hudson1999,keane2001}, from either HCO$+$OH
$\rightarrow$ HCOOH or HCO$+$O $\rightarrow$ HCOO$+$H $\rightarrow$
HCOOH. In addition, experiments suggest that under specific catalytic
conditions CO$_2$ can react to form HCOOH \citep{ogo2006} but this
requires catalytic surface sites, i.e., CO$_2$ directly attached to a
silicate or metallic grain site. Such a situation is less likely in
dense interstellar clouds where thick ice layers have already formed
and cover any potential catalytic sites. In conclusion, under
astrophysically relevant conditions solid CO$_2$ in bulk ice is a very
stable molecule that is not likely to react with H-atoms.

Similar to CO$_2$, reaction rates of HCOOH with H-atoms are below the
detection limit in our experiment. Formation of the so far undetected
interstellar species CH$_2$(OH)$_2$ in this way thus seems
unlikely. Unless other formation mechanisms are found an observational
search for this species based upon solid state astrochemical arguments
is not warranted. We conclude that CO$_2$, HCOOH and CH$_2$(OH)$_2$ do
not appear to be related through successive hydrogenation in
interstellar ice analogues under the conditions as used in the present
study.

In contrast to CO$_2$ and HCOOH, CO does react with H-atoms. The
reaction rates of CO in CO:CO$_2$ mixtures are very similar to those
found by \citet{fuchs2007} for pure CO ices. CO hydrogenation in
interstellar ices will thus not be strongly affected by the presence
of CO$_2$ in the ice. It is likely that the reaction rate is the same
for H$+$CO independent of the CO concentration and that of other
species in apolar interstellar ices.

Reactions of CH$_3$CHO in interstellar ices proceed at similar rates
compared to CO hydrogenation. A maximum of 20\% will be converted to
ethanol, C$_2$H$_5$OH, while another major reaction channel leads to
CH$_4$, H$_2$CO and CH$_3$OH. The precise abundance of CH$_3$CHO in
interstellar ice is not yet well determined. However, abundances of
1--5\% are quoted in the literature
\citep{schutte1997,schutte1999,gibb2004,boogert2004}. These values can
be used to derive an upper limit on the C$_2$H$_5$OH abundance that
could thus be formed. As an example we compare the abundances for the
high mass source W~33A, where CH$_3$CHO has a solid state abundance of
9.8$\times$10$^{-6}$ and CH$_3$OH of 1.4--1.7$\times$10$^{-5}$ both
with respect to H$_2$. If we assume that all solid CH$_3$CHO is
present in the surface layer and the C$_2$H$_5$OH yield is $\sim$20\%,
an abundance of C$_2$H$_5$OH of at most 2.0$\times$10$^{-6}$ with
respect to H$_2$ can be formed. This leads to an upper limit on the
C$_2$H$_5$OH/CH$_3$OH ratio of 0.14. In reality this value will be
lower as part of the CH$_3$CHO ice may be shielded from incoming
H-atoms and other destruction reactions will likely be competing with
hydrogenation reactions. The limit of 0.14 is clearly higher than the
observationally derived C$_2$H$_5$OH/CH$_3$OH abundance ratio in the
gas phase of 0.025$\pm$0.013 \citep{bisschop2007a}. Formation of
C$_2$H$_5$OH from solid state hydrogenation of CH$_3$CHO is thus
sufficient to explain the observed abundances of C$_2$H$_5$OH.

\section{Summary and conclusions}\label{sum}

Hydrogenation reactions of CO$_2$, HCOOH and CH$_3$CHO interstellar
ice analogues have been studied under ultra-high vacuum
conditions. RAIRS and TPD have been used to analyze the results. From
these experiments reaction rates and upper limits on destruction and
formation rates of the above mentioned species are calculated. The
main conclusions derived from this work are:

\begin{itemize}
\item CO$_2$ and HCOOH do not react with H-atoms at a detectable
  level. Only minor fractions of the species desorb due to the
  bombardment. Solid state formation of HCOOH from CO$_2$ and
  CH$_2$(OH)$_2$ from HCOOH are likely inefficient in interstellar
  ices.

\item Hydrogenation of CO to H$_2$CO and CH$_3$OH from CO mixed with
  CO$_2$ has similar reaction rates compared to pure CO ices. The
  presence of CO$_2$ in interstellar ices with CO therefore does not
  affect the formation of H$_2$CO and CH$_3$OH.

\item Hydrogenation of CH$_3$CHO leads for $\sim$20\% to C$_2$H$_5$OH,
  showing for the first time that a thermal hydrogenation reaction can
  be responsible for the C$_2$H$_5$OH abundances detected in dense
  interstellar clouds. Other reaction products are H$_2$CO, CH$_3$OH
  (15--50\%) and CH$_4$ ($\sim$10\%). Due to the energy released a
  fraction of the produced species may evaporate into the gas phase
  upon formation.

\end{itemize}

\begin{acknowledgements}
  Funding was provided by NOVA, the Netherlands Research School for
  Astronomy and by a Spinoza grant from the Netherlands Organization
  for Scientific Research, NWO. We thank Sergio Ioppolo for help with
  the experiments, Herma Cuppen for the theoretical simulations of
  hydrogen flux in our experiment, Karin {\"O}berg for stimulating
  discussions and an anonymous referee for constructive comments on
  the paper.
\end{acknowledgements}


\begin{thebibliography}{47}
\expandafter\ifx\csname natexlab\endcsname\relax\def\natexlab#1{#1}\fi

\bibitem[{{Allamandola} {et~al.}(1988){Allamandola}, {Sandford}, \&
  {Valero}}]{allamandola1988}
{Allamandola}, L.~J., {Sandford}, S.~A., \& {Valero}, G.~J. 1988, Icarus, 76,
  225

\bibitem[{{Andersson} {et~al.}(2006){Andersson}, {Al-Halabi}, {Kroes}, \& {van
  Dishoeck}}]{andersson2006}
{Andersson}, S., {Al-Halabi}, A., {Kroes}, G.-J., \& {van Dishoeck}, E.~F.
  2006, \jcp, 124, 4715

\bibitem[{{Benitez} {et~al.}(1993){Benitez}, {Carrizosa}, \&
  {Odriozola}}]{benitez1993}
{Benitez}, J.~J., {Carrizosa}, J., \& {Odriozola}, J.~A. 1993, Appl. Surf.
  Sci., 68, 565

\bibitem[{{Bennett} {et~al.}(2005){Bennett}, {Jamieson}, {Osamura}, \&
  {Kaiser}}]{bennett2005a}
{Bennett}, C.~J., {Jamieson}, C.~S., {Osamura}, Y., \& {Kaiser}, R.~I. 2005,
  \apj, 624, 1097

\bibitem[{{Bis\-schop} {et~al.}(2007{\natexlab{a}}){Bis\-schop}, {Fuchs},
  {Boogert}, {van Dishoeck}, \& {Linnartz}}]{bisschop2007b}
{Bis\-schop}, S.~E., {Fuchs}, G.~W., {Boogert}, A.~C.~A., {van Dishoeck},
  E.~F., \& {Linnartz}, H. 2007{\natexlab{a}}, \aap, 470, 749

\bibitem[{{Bis\-schop} {et~al.}(2007{\natexlab{b}}){Bis\-schop},
  {J{\o}rgensen}, {van Dishoeck}, \& {de Wachter}}]{bisschop2007a}
{Bis\-schop}, S.~E., {J{\o}rgensen}, J.~K., {van Dishoeck}, E.~F., \& {de
  Wachter}, E.~B.~M. 2007{\natexlab{b}}, \aap, 465, 913

\bibitem[{{Blake} {et~al.}(1987){Blake}, {Sutton}, {Masson}, \&
  {Phillips}}]{blake1987}
{Blake}, G.~A., {Sutton}, E.~C., {Masson}, C.~R., \& {Phillips}, T.~G. 1987,
  \apj, 315, 621

\bibitem[{{Boogert} {et~al.}(2004){Boogert}, {Pontoppidan}, {Lahuis},
  {J{\o}rgensen}, {Augereau}, {Blake}, {Brooke}, {Brown}, {Dullemond}, {Evans},
  {Geers}, {Hogerheijde}, {Kessler-Silacci}, {Knez}, {Morris},
  {Noriega-Crespo}, {Sch{\"o}ier}, {van Dishoeck}, {Allen}, {Harvey},
  {Koerner}, {Mundy}, {Myers}, {Padgett}, {Sargent}, \&
  {Stapelfeldt}}]{boogert2004}
{Boogert}, A.~C.~A., {Pontoppidan}, K.~M., {Lahuis}, F., {et~al.} 2004, \apjs,
  154, 359

\bibitem[{{Bouwman} {et~al.}(2007){Bouwman}, {Ludwig}, {Awad}, {{\"O}berg},
  {Fuchs}, {van Dishoeck}, \& {Linnartz}}]{bouwman2007}
{Bouwman}, J., {Ludwig}, W., {Awad}, Z., {et~al.} 2007, \aap\ in press

\bibitem[{{Cox} {et~al.}(1989){Cox}, {Wagman}, \& {Medvedev}}]{cox1989}
{Cox}, J.~D., {Wagman}, D.~D., \& {Medvedev}, V.~A. 1989, {CODATA Key Values
  for Thermodynamics} (Hemisphere Publishing Corp.)

\bibitem[{{Cyriac} \& {Pradeep}(2005)}]{cyriac2005}
{Cyriac}, J. \& {Pradeep}, T. 2005, Chem. Phys. Lett., 402, 116

\bibitem[{{Ehrenfreund} {et~al.}(1999){Ehrenfreund}, {Kerkhof}, {Schutte},
  {Boogert}, {Gerakines}, {Dartois}, {D'Hendecourt}, {Tielens}, {van Dishoeck},
  \& {Whittet}}]{ehrenfreund1999}
{Ehrenfreund}, P., {Kerkhof}, O., {Schutte}, W.~A., {et~al.} 1999, \aap, 350,
  240

\bibitem[{{Ewing} {et~al.}(1960){Ewing}, {Thompson}, \& {Pimentel}}]{ewing1960}
{Ewing}, G.~E., {Thompson}, W.~E., \& {Pimentel}. 1960, J. Chem. Phys., 32, 927

\bibitem[{{Frenkel} {et~al.}(1994){Frenkel}, {Marsh}, {Wilhoit}, {Kabo}, \&
  {Roganov}}]{frenkel1994}
{Frenkel}, M., {Marsh}, K.~N., {Wilhoit}, R.~C., {Kabo}, G.~J., \& {Roganov},
  G.~N. 1994, {Thermodynamics of Organic Compounds in the Gas State} (Texas,
  U.S.A.: Thermodynamics Research Center)

\bibitem[{{Fuchs} {et~al.}(2007){Fuchs}, {Ioppolo}, {Bisschop}, {Van Dishoeck},
  \& {Linnartz}}]{fuchs2007}
{Fuchs}, G.~W., {Ioppolo}, S., {Bisschop}, S.~E., {Van Dishoeck}, E.~F., \&
  {Linnartz}, H. 2007, submitted to \aap

\bibitem[{{Gerakines} {et~al.}(2000){Gerakines}, {Moore}, \&
  {Hudson}}]{gerakines2000}
{Gerakines}, P.~A., {Moore}, M.~H., \& {Hudson}, R.~L. 2000, \aap, 357, 793

\bibitem[{{Gibb} {et~al.}(2004){Gibb}, {Whittet}, {Boogert}, \&
  {Tielens}}]{gibb2004}
{Gibb}, E.~L., {Whittet}, D.~C.~B., {Boogert}, A.~C.~A., \& {Tielens},
  A.~G.~G.~M. 2004, \apjs, 151, 35

\bibitem[{{Gurvich} {et~al.}(1989){Gurvich}, {Veyts}, \&
  {Alcock}}]{gurvich1989}
{Gurvich}, L.~V., {Veyts}, I.~V., \& {Alcock}, C.~B. 1989, {Thermodynamic
  Properties of Individual Substances}, 4th edn. (New York: Hemisphere Pub.
  Co.)

\bibitem[{{Hidaka} {et~al.}(2004){Hidaka}, {Watanabe}, {Shiraki}, {Nagaoka}, \&
  {Kouchi}}]{hidaka2004}
{Hidaka}, H., {Watanabe}, N., {Shiraki}, T., {Nagaoka}, A., \& {Kouchi}, A.
  2004, \apj, 614, 1124

\bibitem[{{Hiraoka} {et~al.}(1994){Hiraoka}, {Ohashi}, {Kihara}, {Yamamoto},
  {Sato}, \& {Yamashita}}]{hiraoka1994}
{Hiraoka}, K., {Ohashi}, N., {Kihara}, Y., {et~al.} 1994, Chem. Phys. Lett.,
  229, 408

\bibitem[{{Hiraoka} {et~al.}(2002){Hiraoka}, {Sato}, {Sato}, {Sogoshi},
  {Yokoyama}, {Takashima}, \& {Kitagawa}}]{hiraoka2002}
{Hiraoka}, K., {Sato}, T., {Sato}, S., {et~al.} 2002, \apj, 577, 265

\bibitem[{{Hudson} \& {Moore}(1999)}]{hudson1999}
{Hudson}, R.~L. \& {Moore}, M.~H. 1999, Icarus, 140, 451

\bibitem[{{Hwang} \& {Mebel}(2004)}]{hwang2004}
{Hwang}, d.~Y. \& {Mebel}, A.~M. 2004, J. Phys. Chem. A, 108, 10245

\bibitem[{{Ikeda} {et~al.}(2001){Ikeda}, {Ohishi}, {Nummelin}, {Dickens},
  {Bergman}, {Hjalmarson}, \& {Irvine}}]{ikeda2001}
{Ikeda}, M., {Ohishi}, M., {Nummelin}, A., {et~al.} 2001, \apj, 560, 792

\bibitem[{{Keane}(2001)}]{keane2001}
{Keane}, J.~V. 2001, PhD thesis, Rijks Universiteit Groningen

\bibitem[{{Keane} {et~al.}(2001){Keane}, {Tielens}, {Boogert}, {Schutte}, \&
  {Whittet}}]{keane2001a}
{Keane}, J.~V., {Tielens}, A.~G.~G.~M., {Boogert}, A.~C.~A., {Schutte}, W.~A.,
  \& {Whittet}, D.~C.~B. 2001, \aap, 376, 254

\bibitem[{{Lakin} {et~al.}(2003){Lakin}, {Troya}, {Schatz}, \&
  {Harding}}]{lakin2003}
{Lakin}, M.~J., {Troya}, D., {Schatz}, G.~C., \& {Harding}, L.~B. 2003, J.
  Chem. Phys., 119, 5848

\bibitem[{{Lugez} {et~al.}(1994){Lugez}, {Schriver}, {Levant}, \&
  {Schriver-Mazzuoli}}]{lugez1994}
{Lugez}, C., {Schriver}, A., {Levant}, R., \& {Schriver-Mazzuoli}, L. 1994,
  Chem. Phys., 181, 129

\bibitem[{{Matus} {et~al.}(2007){Matus}, {Nguyen}, \& {Dixon}}]{matus2007}
{Matus}, M.~H., {Nguyen}, M.~T., \& {Dixon}, D.~H. 2007, J. Phys. Chem. A, 111,
  113

\bibitem[{{Milligan} \& {Jacox}(1964)}]{milligan1964}
{Milligan}, D.~E. \& {Jacox}, M.~E. 1964, J. Chem. Phys., 41, 3032

\bibitem[{{Milligan} \& {Jacox}(1971)}]{milligan1971}
{Milligan}, D.~E. \& {Jacox}, M.~E. 1971, J. Chem. Phys., 54, 927

\bibitem[{{Moore} \& {Hudson}(1998)}]{moore1998}
{Moore}, M.~H. \& {Hudson}, R.~L. 1998, Icarus, 135, 518

\bibitem[{{Moore} \& {Hudson}(2003)}]{moore2003}
{Moore}, M.~H. \& {Hudson}, R.~L. 2003, Icarus, 161, 486

\bibitem[{{Moore} \& {Hudson}(2005)}]{moore2005}
{Moore}, M.~H. \& {Hudson}, R.~L. 2005, in IAU Symposium, Vol. 231,
  Astrochemistry: Recent Successes and Current Challenges, ed. D.~C. {Lis},
  G.~A. {Blake}, \& E.~{Herbst}, 247--260

\bibitem[{{Moore} {et~al.}(2001){Moore}, {Hudson}, \& {Gerakines}}]{moore2001}
{Moore}, M.~H., {Hudson}, R.~L., \& {Gerakines}, P.~A. 2001, Spectrochim. Acta
  A, 57, 843

\bibitem[{{{\"O}berg} {et~al.}(2007){{\"O}berg}, {Fraser}, {Boogert},
  {Bisschop}, {Fuchs}, {van Dishoeck}, \& {Linnartz}}]{oberg2007}
{{\"O}berg}, K.~I., {Fraser}, H.~J., {Boogert}, A.~C.~A., {et~al.} 2007, \aap,
  462, 1187

\bibitem[{{Ogo} {et~al.}(2006){Ogo}, H., {Harada}, \& {Fukuzumi}}]{ogo2006}
{Ogo}, S.and~{Kabe}, R., H., H., {Harada}, R., \& {Fukuzumi}, S. 2006, Dalton
  T., 39, 4657

\bibitem[{{Ruscic} {et~al.}(2002){Ruscic}, {Wagner}, {Harding}, {Asher},
  {Feller}, {Dixon}, {Peterson}, {Song}, {Qian}, {Ng}, {Liu}, \&
  {Chen}}]{ruscic2002}
{Ruscic}, B., {Wagner}, A.~F., {Harding}, L.~B., {et~al.} 2002, J. Chem. Phys.
  A, 106, 2727

\bibitem[{{Schutte} {et~al.}(1999){Schutte}, {Boogert}, {Tielens}, {Whittet},
  {Gerakines}, {Chiar}, {Ehrenfreund}, {Greenberg}, {van Dishoeck}, \& {de
  Graauw}}]{schutte1999}
{Schutte}, W.~A., {Boogert}, A.~C.~A., {Tielens}, A.~G.~G.~M., {et~al.} 1999,
  \aap, 343, 966

\bibitem[{{Schutte} {et~al.}(1997){Schutte}, {Greenberg}, {van Dishoeck},
  {Tielens}, {Boogert}, \& {Whittet}}]{schutte1997}
{Schutte}, W.~A., {Greenberg}, J.~M., {van Dishoeck}, E.~F., {et~al.} 1997,
  \apss, 255, 61

\bibitem[{{Tielens} \& {Charnley}(1997)}]{tielens1997}
{Tielens}, A.~G.~G.~M. \& {Charnley}, S.~B. 1997, Origins Life Evol. B., 27, 23

\bibitem[{{Tschersich}(2000)}]{tschersich2000}
{Tschersich}, K.~G. 2000, Journal of Applied Physics, 87, 2565

\bibitem[{{Tschersich} \& {von Bonin}(1998)}]{tschersich1998}
{Tschersich}, K.~G. \& {von Bonin}, V. 1998, Journal of Applied Physics, 84,
  4065

\bibitem[{{Van Ijzendoorn} {et~al.}(1983){Van Ijzendoorn}, {Allamandola},
  {Baas}, \& {Greenberg}}]{ijzendoorn1983}
{Van Ijzendoorn}, L.~J., {Allamandola}, L.~J., {Baas}, F., \& {Greenberg},
  J.~M. 1983, J. Chem. Phys., 78, 7019

\bibitem[{{Watanabe} {et~al.}(2004){Watanabe}, {Nagaoka}, {Shiraki}, \&
  {Kouchi}}]{watanabe2004}
{Watanabe}, N., {Nagaoka}, A., {Shiraki}, T., \& {Kouchi}, A. 2004, \apj, 616,
  638

\bibitem[{{Wiberg} {et~al.}(1991){Wiberg}, {Crocker}, \& {Morgan}}]{wiberg1991}
{Wiberg}, K.~B., {Crocker}, L.~S., \& {Morgan}, K.~M. 1991, J. Am. Chem. Soc,
  113, 3447

\bibitem[{{Wu} {et~al.}(2002){Wu}, {Judge}, {Cheng}, {Shih}, {Yih}, \&
  {Ip}}]{wu2002}
{Wu}, C.~Y.~R., {Judge}, D.~L., {Cheng}, B.-M., {et~al.} 2002, Icarus, 156, 456

\end{thebibliography}
\end{document}